\documentclass[doublecol,figures]{epl2}%
\usepackage{epsfig}
\usepackage{caption}
\usepackage{lscape}
\usepackage{bm}
\usepackage{amsfonts}
\usepackage{amsmath}
\usepackage{amssymb}
\usepackage{graphicx}%
\institute{
\inst{1} Department of Physics, and Center of Theoretical and Computational Physics,
The University of Hong Kong, Hong Kong, China\\
\inst{2} Department of Physics, Renmin University of China, Beijing 100872, China\\
\inst{3} Beijing National Laboratory for Condensed Matter Physics and Institute of
Physics, Chinese Academy of Sciences, Beijing 100080, China
}
\pacs{72.25.-b}{Spin polarized transport}
\pacs{73.21.La}{Quantum dots}
\pacs{72.15.Qm}{Scattering mechanisms and Kondo effect}
\abstract{Quantum spin transport is studied in an interacting quantum dot. It is found
that a conductance "plateau" emerges in the non-linear charge conductance by a
spin bias in the Kondo regime. The conductance plateau, as a complementary to
the Kondo peak, originates from the strong electron correlation and exchange
processes in the quantum dot, and can be regarded as one of the
characteristics in quantum spin transport.
}
\begin{document}

\title{Conductance plateau in quantum spin transport through an interacting quantum dot}
\author{Y. J. Bao\inst{1}
\and N. H. Tong\inst{2}
\and Q. F. Sun\inst{3}
\and S. Q. Shen\inst{1}}
\maketitle

\section{Introduction}

The quantum charge transport has been extensively investigated in a wide class
of correlated electron systems, which involves the interaction between a
localized spin and free conduction electrons, such as in metals containing
magnetic impurities \cite{Rev}. Among them, the Kondo effect was of
fundamental importance and was proposed to emerge in the device consisting of
a single-electron quantum dot (QD) coupled to electrodes (i.e. leads)
\cite{theor1,theor2}, and its conductance can be strongly enhanced in the
Kondo regime due to the frequent occurrence of spin exchange processes. Rapid
progresses in nano-technology have made it available to control the number of
electrons in QD and the parameters (e.g. the energy level, coupling strength
between QD and leads, and so on) of the QD \cite{exper1}. The Kondo effect in
the QD was observed experimentally \cite{experK1,experK2}, which was in
excellent agreement with theoretical predictions \cite{theor1, theor2}. Most
early works focus on the Kondo physics driven by the charge current, which was
generated in two non-magnetic leads and under the bias voltage $V$ (hereafter
the bias voltage is named as charge bias). Recently the Kondo effect of the
spin-polarized transport in QD with the ferromagnetic leads was investigated
\cite{Ferro1,Ferro2}. It reveals novel properties of the Kondo effect in the
spin aspect. Since electron has both charge and spin, it would be imcomplete
for any physical picture of the Kondo effect without full consideration of
quantum spin transport. On the other hand, the spin current, in which
electrons with spin up and down move in opposite directions, has attracted a
lot of theoretical and experimental attentions. By using the magnetic
tunneling injection, and electrical or optical injection technique, a pure
spin current without accompanying the charge current can be generated and
detected experimentally \cite{experH1,experH2}. These experimental progresses
make it practical to investigate the properties of quantum spin transport in
correlated systems.

In the present paper, we introduce a \textrm{spin bias} instead of a
\textrm{charge bias} to study the quantum transport in Kondo regime in an
interacting QD system. The spin bias means the spin-dependent chemical
potential in the two leads, and may produce a spin current flowing through the
QD (see Fig.2) \cite{sbias}. Consider a common-used device, which consists of
a QD coupled to two non-magnetic leads. It is found that a conductance plateau
emerges in the non-linear charge conductance while a double peak appears in
the non-linear spin conductance. The charge conductance plateau and spin
conductance double peak only exist in the Kondo regime, and its width is equal
to twice of the spin splitting of the QD's energy levels. It is believed that
these phenomena can be regarded as one of the intrinsic characteristics in
quantum spin transport.

\section{General formalism}

\subsection{Model Hamiltonian and the spin bias}

We start with the model Hamiltonian for the QD coupled to two leads
\cite{theor2},
\begin{equation}
H=H_{Lead}+H_{D}+H_{T}\,.
\end{equation}
Here
\begin{equation}
H_{Lead}={\sum\limits_{\alpha,k,\sigma}}\epsilon_{\alpha k}c_{\alpha k\sigma
}^{\dag}c_{\alpha k\sigma}%
\end{equation}
describes the conduction electrons in the left (L) and right (R) leads
$\alpha\in(L,R)$ and $c_{\alpha k\sigma}^{\dag}\left(  c_{\alpha k\sigma
}\right)  $ is the creation (annihilation) operator for electrons with
momentum $k$ and spin $\sigma\in(\uparrow,\downarrow)$ in the lead $\alpha$.
\begin{equation}
H_{D}={\sum\limits_{\sigma}}\epsilon_{\sigma}n_{\sigma}+Un_{\uparrow
}n_{\downarrow}(n_{\sigma}=d_{\sigma}^{\dag}d_{\sigma})
\end{equation}
is the Hamiltonian for the interacting QD and, for simplicity, we only
consider a single pair of levels with spin splitting $\epsilon_{\sigma
}=\epsilon_{d}+\sigma\triangle/2$. $U$ represents the Coulomb interaction.
\begin{equation}
H_{T}={\sum\limits_{\alpha,k,\sigma}}\left(  t_{\alpha k}c_{\alpha k\sigma
}^{\dag}d_{\sigma}+h.c.\right)
\end{equation}
describes the tunneling between the QD and two leads with the tunneling
coefficients $t_{\alpha k}$. The Kondo effect in this device has been
extensively studied \cite{theor1,theor2}. Most previous works focused on the
case of the charge bias $V$ between two leads with the chemical potential
$\mu_{L\sigma}=-\mu_{R\sigma}=V/2$, and part of the works were also concerned
with the spin-polarized current driven Kondo effect \cite{Ferro1,Ferro2}. In
the following, we apply the spin bias $V$, where
\begin{align}
\mu_{L\uparrow}  &  =-\mu_{R\downarrow}=V/2,\\
\mu_{L\downarrow}  &  =-\mu_{R\uparrow}=-V/2\nonumber
\end{align}
\ (see Fig. 2) \cite{sbias} to study the quantum electron transport. This spin
bias can generate a spin current, in which the electrons with different spins
move in opposite directions.

\subsection{Theory of Keldysh non-equilibrium Green's functions technique}

In the theory of the Keldysh non-equilibrium Green's functions technique, the
current from the lead $\alpha$ flowing into the QD for the spin channel
$\sigma$ can be expressed as \cite{curr},
\begin{equation}
I_{\alpha\sigma}=\frac{i}{\hbar}\int\frac{d\,\epsilon}{2\pi}\Gamma_{\alpha
}(\epsilon)\left\{  f_{\alpha\sigma}(\epsilon)\left(  G_{\sigma}^{r}%
(\epsilon)-G_{\sigma}^{a}(\epsilon)\right)  +G_{\sigma}^{<}(\epsilon)\right\}
, \label{current}%
\end{equation}
where $\Gamma_{\alpha}(\epsilon)=2\pi\sum_{k}t_{\alpha k}^{\ast}t_{\alpha
k}\,\delta(\epsilon-\epsilon_{\alpha k})$ is the line-width function and
$f_{\alpha\sigma}(\epsilon)=\left\{  \exp\left[  (\epsilon-\mu_{\alpha\sigma
})/k_{B}T\right]  +1\right\}  ^{-1}$ is the Fermi-Dirac distribution of
electrons in the lead $\alpha$ with spin $\sigma$ at temperature $T$.
$G_{\sigma}^{r,a,<}(\epsilon)$ are the Fourier transformation of the retarded,
advanced and lesser Green's function $G_{\sigma}^{r,a,<}(t)$ in the QD,
respectively, where $G_{\sigma}^{<}(t)\equiv i\left\langle d_{\sigma}^{\dag
}(0)\,d_{\sigma}(t)\right\rangle \,$ and $G_{\sigma}^{r,a}(t)\equiv\mp
i\theta(\pm t)\left\langle \left\{  d_{\sigma}(t)\,,d_{\sigma}^{\dag
}(0)\right\}  \right\rangle \,$.

The Green's functions can be solved by the equation-of-motion technique, which
was widely applied to study the electron transport in interacting QDs
\cite{theor2,book}. Although this method has some disadvantages in predicting
the intensity of Kondo effect and the disappearance of the Kondo peak at
$\epsilon_{d}+2U=0$, it is relatively simple and straightforward, and has been
proven to provide correct \textit{qualitative} physics at low temperatures and
in the large $U$ limit ($U\rightarrow\infty$). For the sake of simplicity, we
concentrate on the transport properties in the strong interaction limit, i.e.,
$U\rightarrow\infty$. In this case, the Fourier transform of the Green's
functions $G_{\sigma}^{r,a}(\epsilon)$ is obtained as,
\begin{equation}
G_{\sigma}^{r,a}(\epsilon)=\frac{1-n_{\bar{\sigma}}}{\epsilon-\epsilon
_{\sigma}-\Sigma_{0\sigma}^{r,a}(\epsilon)-\Sigma_{1\sigma}^{r,a}(\epsilon
)}\,,
\end{equation}
where $n_{\sigma}$ is the occupation number in the spin-$\sigma$ state and
$\Sigma_{0\sigma}^{r,a}(\epsilon)=2\pi\sum\limits_{\alpha k}|t_{\alpha k}%
|^{2}/(\epsilon-\epsilon_{\alpha k}\pm i0^{+})$ is the non-interacting
tunneling self-energy. $\Sigma_{0\sigma}^{r,a}(\epsilon)$ can reduce to
$\Sigma_{0\sigma}^{r,a}(\epsilon)=\mp i(\Gamma_{L}+\Gamma_{R})/2$ while in the
wide-band width limit, i.e. $\Gamma_{L}$ and $\Gamma_{R}$ being a constant
independent of $\epsilon$. $\Sigma_{1\sigma}^{r,a}(\epsilon)$ in Eq.(7) due to
the electron correlation and the tunneling has the form:
\begin{equation}
\Sigma_{1\sigma}^{r,a}(\epsilon)=\sum_{\alpha k}\frac{f_{\alpha\sigma
}(\epsilon_{\alpha k})\left\vert t_{\alpha k}\right\vert ^{2}}{\epsilon
-\epsilon_{\alpha k}+\epsilon_{\bar{\sigma}}-\epsilon_{\sigma}\pm i0^{+}}\,.
\end{equation}
The occupation number $n_{\sigma}$ in Eq.(7) should be solved
self-consistently from the equation, $n_{\sigma}=-i\int(d\epsilon
/2\pi)G_{\sigma}^{<}(\epsilon)$.

Next we come to solve the lesser Green's function $G_{\sigma}^{<}(\epsilon)$
to calculate the current formula in Eq.(6) and the occupation number
$n_{\sigma}$ self-consistently. For interacting systems, usually $G_{\sigma
}^{<}(\epsilon)$ cannot be solved exactly, and various approximations were
developed such as Ng ansatz \cite{Ng}, and the non-crossing approximation
\cite{theor2}. In the present calculation, instead of $G_{\sigma}^{<}%
(\epsilon)$ only $\int d\epsilon G_{\sigma}^{<}(\epsilon)$ is actually used
\cite{sun},
\begin{equation}
\int d\epsilon G_{\sigma}^{<}(\epsilon)=-i\int d\epsilon\text{Im}G_{\sigma
}^{r}(\epsilon)\left(  f_{L\sigma}(\epsilon)+f_{R\sigma}(\epsilon)\right)  \,.
\end{equation}
With the help of Eqs.(7), (8), and (9), we are ready to calculate the particle
current $I_{\alpha\sigma}$. Since there is no spin-flip process in the system,
we have $dn_{\sigma}/dt=-(I_{R\sigma}+I_{L\sigma})$ and $I_{L\sigma
}=-I_{R\sigma}\equiv I_{\sigma}$ in a steady state. As a result, the charge
current $I_{c}=e(I_{\uparrow}+I_{\downarrow})$ and the spin current
$I_{s}=\frac{\hbar}{2}(I_{\uparrow}-I_{\downarrow})$ are conserved. In the
following calculations, the line-width function $\Gamma_{\alpha}(\epsilon)$ is
taken to be symmetric $\Gamma_{L}(\epsilon)=\Gamma_{R}(\epsilon)=\Gamma
\theta(D-\left\vert \epsilon\right\vert ),$ with a band width $2D=1000$, and
$\Gamma=1$ as the unit of energy. We define the charge differential
conductance
\begin{equation}
G_{c}(V)=dI_{c}/dV\,,
\end{equation}
and the spin differential conductance
\begin{equation}
G_{s}(V)=dI_{s}/dV\,,
\end{equation}
\ with respect to the spin bias $V$. In the absence of the Zeeman splitting
$\Delta=0$, $I_{\uparrow}=-I_{\downarrow}$ because the Hamiltonian is
invariant under the exchange of the indices ($L,\uparrow$) and ($R,\downarrow
$). In this case the charge current $I_{c}$ and its conductance $\sigma_{c}$
are equal to zero exactly. To obtain a non-zero $I_{c}$ and $\sigma_{c}$ the
energy levels of the QD must be split, i.e. $\Delta\not =0$. For the spin
current $I_{s}$ and its conductance $\sigma_{s}$ they are always non-zero
under a spin bias even if the energy levels are degenerate.

\begin{figure}[ptb]
\centering \includegraphics[width=0.45\textwidth]{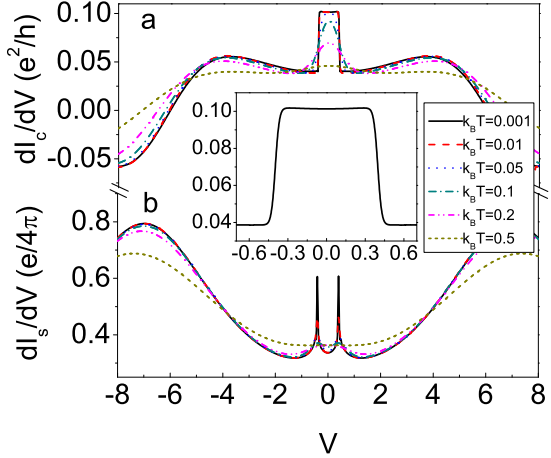}\newline%
\caption{ (color online). The charge conductance $dI_{c}/dV$ in (a) and the
spin conductance $dI_{s}/dV$ in (b) vs. the spin bias $V$ at different
temperatures $T$. The inset magnifies the curve of $dI_{c}/dV$ for
$k_{B}T=0.01$. ($\Gamma=1$, $\epsilon_{d}=-5$, $\triangle=0.4$.)}%
\label{GV-V-Tdiff}%
\end{figure}%

It is worth pointing out that it is also possible to generate a
charge current as well as a spin current
if the two tunneling coefficients are unequal, $t_{L,k}\not= t_{R,k}$,
even in the absence of the Zeeman splitting. The right-left asymmetry plus the
strong Coulomb interaction may generate a charge current. The self-energy correction
will also produce the splitting of the energy level in the quantum dot, which
looks like an  effective "Zeeman" energy splitting, and varies with the strength of
spin bias. However, the splitting is quite tiny.

\section{Results and discussions}

The main results in the present paper are summarized in Fig. 1(a) where a
conductance plateau appears in the curve of the charge conductance $dI_{c}/dV$
and in Fig. 1(b) where a double peak appears in the curve of the spin
conducatnce $dI_{s}/dV$ with respect to the spin bias $V$ at different
temperatures $T$. The double-peak occurs at $V=\pm\Delta$ due to the energy
splitting by a magnetic field. A similar result was observed from a solvable model.\cite{Katsura07}
It is also very similar to the Kondo double-peak which is induced by the
charge bias $V_{c}$ in the curve of the conventional charge conductance
$dI/dV_{c}$-$V_{c}$, and has been observed experimentally
\cite{Rev,experK1,experK2}. The relationship between them will be discussed in
more detail elsewhere. In the following, we focus mainly on the charge
conductance induced by the spin bias. In Fig. 1(a), a plateau instead of the
conventional Kondo-double-peak appears between $-\Delta<V<\Delta,$ which is
the key feature of spin-bias-induced conductance $dI_{c}/dV$. This plateau is
quite even, and the relative variance $|G_{c}(V)-G_{c}(0)|/G_{c}(0)$ is within
0.1\% in a quite large region of the spin bias voltage $V$ (about
$20k_{B}T_{K}$ at $\Delta=0.4)$ \cite{note1}. At $V=\pm\Delta$, the
conductance $dI_{c}/dV$ suddenly drops down within the scale of $k_{B}T_{K}$
\footnote{Although the Kondo temperature $T_{K}=D\exp[-\pi\epsilon_{d}%
/(\Gamma_{L}+\Gamma_{R})]\approx0.2$ for $\epsilon_{d}=-5\Gamma$, $T_{K}$ is
estimated to be about $0.04$ from the half-width of the Kondo peak in Fig. 1b.
Usually the equation-of-motion method underestimates the electron correlation
and gives a lower value of $T_{K}$. For the discussion in the text, we take
$k_{B}T_{K}=0.04$ for $\epsilon_{d}=-5\Gamma$.}. In particular, this plateau
has the same properties as the Kondo peak: it disappears completely if $U=0$,
or at higher temperatures $T>T_{K}$. The plateau survives only at low
temperatures ($T\lesssim T_{K}$), and tends to saturate if $T\ll T_{K}$ (see
Fig. 1(a)).

\begin{figure}[ptbh]
\centering \includegraphics[width=0.45\textwidth]{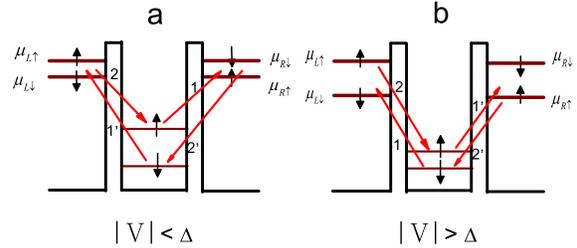}\newline\caption{
(color online). (a) and (b) are the schematic diagrams for the dominant
exchange processes in $|V|<\Delta$ and $|V|>\Delta$, respectively. }%
\end{figure}

The conductance plateau is believed to originate from strong electron
correlation and the exchange process as shown in Fig. 2. The strong Coulomb
interaction only allows that the two levels in the QD can be occupied by at
most a single electron. When $\left\vert V\right\vert <\Delta$ as shown in
Fig. 2(a), there exist two exchange processes: (i) the spin-up electron in the
QD tunnels into the right lead (\textit{the step 1}) and another spin-up
electron in the left lead tunnels into the QD (\textit{the step 2}) due to
$\mu_{L\uparrow}>\mu_{R\uparrow}$, and (ii) the spin-down electron in the QD
tunnels into the left lead (\textit{the step 1'}) and another spin-down
electron in the right lead tunnels into the QD (\textit{the step 2'}) due to
$\mu_{R\downarrow}>\mu_{L\downarrow}$. These two exchange processes may induce
the charge current but with opposite flowing directions. Also because of
$\epsilon_{d\uparrow}>\epsilon_{d\downarrow}$ at the positive Zeeman splitting
$\Delta$, the charge current from the exchange process (i) is larger than that
from the process (ii), so that the conductance is positive and almost constant
at $\left\vert V\right\vert <\Delta$. On the other hand, while $\left\vert
V\right\vert >\Delta$, the spin exchange processes as shown in Fig. 2(b) also
occur. However, these exchange processes are prohibited when $\left\vert
V\right\vert <\Delta$, because the energy is required to be conserved after
the two virtual tunneling steps 1 and 2 , and 1' and 2' in Fig. 2(b). In
particular, these exchange processes can frequently occur at low temperatures,
so that the exchange processes in Fig. 2(a) are restrained. As a result the
conductance $dI_{c}/dV$ sharply drops while $|V|$ passing through $\Delta$. As
for the weak interaction limit, $U=0$, the QD can be occupied by two electrons
simultaneously, and the exchange processes in Fig. 2(b) do not occur because
of the Pauli exclusion principle, and the conductance pleatau disappears. The
co-tunneling processes in Fig. 2(b) also explain the occurrence of the
double-peak in the spin conductance $dI_{s}/dV$. These processes in step 1 and
2 , and 1' and 2' exchange the spins in the two spin channels in the leads. At
each process of step 1 and 2 , and 1' and 2' it will produce spin current
instead of charge current. Thus a resonance of spin differential condctance
emerges at the critical points $V=\pm\Delta$ and the double-peak arises correspondingly.

\begin{figure}[ptbh]
\centering\includegraphics[width=0.45\textwidth]{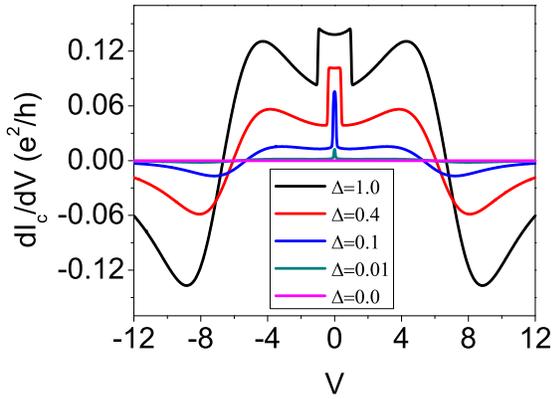}\newline%
\caption{ (color online). The charge conductance $dI_{c}/dV$ vs. the spin bias
$V$ for different $\triangle$. ($k_{B}T=0.01$ and $\epsilon_{d}=-5$.)}%
\end{figure}

To explore the physical properties of the plateau in charge conductance, we
come to study the variance of the plateau with $\Delta$ and $\epsilon_{d}$.
Fig. 3 shows the conductance $dI_{c}/dV$ versus the spin bias $V$ for
different $\Delta$ at $T<T_{K}$ and $\epsilon_{d}=-5$. When $\Delta=0$,
$dI_{c}/dV=0$ for any $V$ and no plateau appears. With increasing $\Delta$
from zero, $\left.  dI_{c}/dV\right\vert _{V=0}$ rises very quickly. The
plateau begins to form when $\Delta$ is about a few $k_{B}T_{K}$. Its width is
$2\Delta$ and its height depends on $\Delta$ weakly. This plateau remains
robust even when $\Delta$ reaches about $20k_{B}T_{K}$. At a larger $\Delta$,
a slightly downward bend emerges in the center of the plateau.

\begin{figure}[ptbh]
\centering\includegraphics[width=0.45\textwidth]{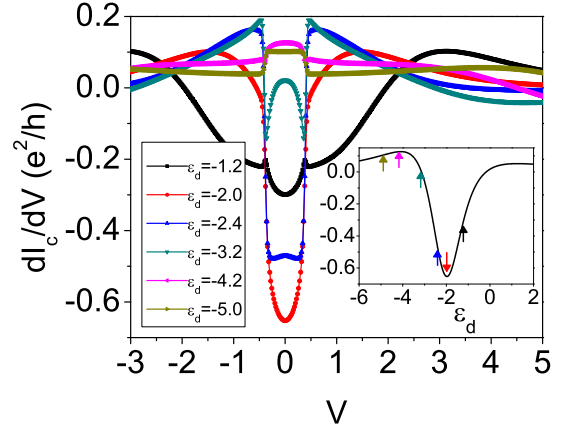}\newline%
\caption{ (color online). The charge conductance $dI_{c}/dV$ vs. the spin bias
$V$ for different $\epsilon_{d}$. The inset plots the curve $dI_{c}%
/dV$-$\epsilon_{d}$ at $V=0$. The arrows indicate the values of $\epsilon_{d}$
in the main figure. ($\Delta=0.4$ and $k_{B}T=0.01$.) }%
\end{figure}

Fig. 4 shows the $\epsilon_{d}$-dependence of the conductance $dI_{c}/dV$
versus the spin bias $V$ at $T<T_{K}$. The conductance plateau always exists
and is robust when $\epsilon_{d}<-5$, i.e. in the Kondo regime where the
electron is localized in the QD. With increasing $\epsilon_{d}$, the electron
is partially delocalized leading to the plateau distorted. When the system is
in the mixed-valence regime, i.e., $-5<\epsilon_{d}\lesssim0$, the plateau
disappears completely, but the two sharp jumps still survive at $V=\pm\Delta$,
and a deep valley emerges at $V=0$. In fact, the linear charge conductance
$G_{c}(0)$ is intensively dependent on $\epsilon_{d}$ in the mixed-valence
regime. The inset of Fig. 4 plots $G_{c}(0)$ versus $\epsilon_{d}$, in which a
deep dip exhibits at about $\epsilon_{d}=-2$ and $G_{c}(0)$ can even change
its sign. At last, when $\epsilon_{d}>0$, the QD is in an empty state, and the
$dI_{c}/dV$-$V$ curve becomes relatively smooth due to the nonexistence of the
exchange process at $\epsilon_{d}>0$. This behavior suggests that the conductance plateau in the Kondo regime is closely related to the localization of electron in the Kondo regime, and the physics behind the plateau deserves further discussion. 

\begin{figure}[ptbh]
\centering \includegraphics[width=0.45\textwidth]{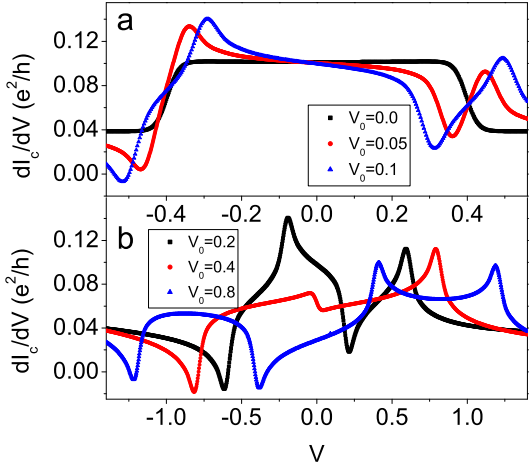}\newline%
\caption{(color online). (a) and (b) show the curve $dI_{c}/dV$-$V$ while both
spin and charge bias coexist. ($\epsilon_{d}=-5$, $\triangle=0.4$, and
$k_{B}T=0.01$.)}%
\end{figure}

Finally we come to discuss the combined effect induced by the spin and charge
bias. Consider the spin bias $V_{s}=V$ and the charge bias $V_{c}=V_{0}$,
\textit{i.e.}, $\mu_{L\uparrow/\downarrow}=\pm V/2+V_{0}$ and $\mu
_{R\uparrow/\downarrow}=\mp V/2$. In this case both spin and charge current
through the QD are non-zero. Fig. 5 plots the $dI_{c}/dV$-$V$ curve for
different charge bias voltages $V_{0}$. The results exhibit very well the
combination of the plateau and the peak. The peaks are of Kondo type induced
by charge bias. For a small $V_{0}$ (see Fig. 5(a)), in addition to the
plateau, two extra Kondo peaks and dips begin to emerge in the $dI_{c}/dV$-$V$
curve at the positions $V=\pm\Delta\pm V_{0}$. The Kondo dips originate from
the fact that the voltage difference $\mu_{L\downarrow}-\mu_{R\downarrow}$
decreases linearly with increasing of the charge bias $V_{0}$, and the
conventional Kondo peaks from the spin-down electron are transformed into the
dips. On the other hand, for a large $V_{0}$ (see Fig. 5(b)), the Kondo peaks
and dips become dominant.

\section{Summary}

In summary, a conductance plateau emerges in the charge differential
conductance $dI_{c}/dV$-$V$ curve in quantum spin transport, instead of the
Kondo-double peak in quantum charge transport. The conductance plateau and
Kondo peak are complementary to each other, and reflect the spin and charge
aspects of the exchange processes in quantum electron transport.

\acknowledgments This work was supported by the Research Grant Council of Hong
Kong under Grant No.: HKU 7041/07P (S.Q.S.); and NSF-China under Grant Nos.
10474125 and 10525418 (Q.F.S.).

\end{document}